\documentclass{pasj00} 
\begin{document}
\SetRunningHead{Y. Sofue}{Galactic Rotation Curve, Dark Halo, and Local Group}
\Received{2008/mm/dd} \Accepted{2008/mm/dd} 

\def\kms{km s$^{-1}$} \def\Msun{M_\odot} 
\def\meleven{\times10^{11}\Msun} \def\mtwelve{\times10^{12}\Msun} 
\def\be{\begin{equation}} \def\ee{\end{equation}}
\def\bc{\begin{center}} \def\ec{\end{center}}

\title{Pseudo Rotation Curve connecting the Galaxy, Dark Halo, and Local Group}
\author{Yoshiaki {\sc Sofue} } 
\affil{Dept. Physics and Astronomy, Kagoshima University, Kagoshima, 890-0065\\ and \\
Institute of Astronomy, University of Tokyo, Mitaka, 181-0015 Tokyo \\
Email:{\it sofue@sci.kagoshima-u.ac.jp and sofue@ioa.s.u-tokyo.ac.jp}
 }


\maketitle

\begin{abstract} 
We construct a Galacto-Local Group rotation curve, combining the Galactic rotation curve with a diagram, where galacto-centric radial velocities of outer globular clusters and member galaxies of the Local Group are plotted against their galacto-centric distances. The high-velocity ends of this pseudo rotation curve within a radius $R\sim150$ kpc are well traced by a rotation curve calculated for the NFW (Navaro, Frenk, White) and Burkert dark halo models. The NFW model indicates that the Galaxy's mass within 385 kpc, half distance to M31, is $\sim 4\meleven$. High-velocity ends of the pseudo rotation curve for the entire Local Group up to 1.5 Mpc indicate isothermal nature with a terminal velocity of $\sim 200$ \kms. In order for the Local Group to be gravitationally bound, an order of magnitude larger mass than those of the Galaxy and M31 is required. This fact suggest that the Local Group contains dark matter of mass $\sim 5\mtwelve$, filling the space between the Galaxy and M31. The mass density of the Galactic dark halo becomes equal to that of the Local Group's dark matter at $R \sim 100$ kpc, beyond which the intracluster dark matter dominates. If we define the Galaxy's radius at this distance, the enclosed Galactic mass is $\sim 3 \meleven$. We comment on the barionic mass fraction of the galaxies in the Local Group.

\end{abstract}

\section{Introduction}

The Milky Way Galaxy  and M31 are the major constituents of the Local Group, around which many satellites and dwarf galaxies are orbiting(Mateo 1998; Sawa and Fujimoto 2005). Masses and extents of dark halos around these two giant galaxies are crucial for understanding the dynamics and structure of the Local Group. Outer rotation curve of the Galaxy is a key to put constraints on the dark halo structure. In this paper, we consider the dark halo of our Galaxy based on a recent accurate rotation curve (Sofue et al. 2008), and try to link it to the dynamics of the Local Group and M31.

Rotation curve of the Galaxy has been extensively used to derive the mass structure (Fich and Tremaine 1991; Sofue and Rubin 2001). The inner rotation curve was obtained by the terminal velocity method applied to radio line observations (Burton, Gordon 1978; Clemens 1985; Fich et al. 1989). The outer rotation curve has been obtained by combining CO line velocities and optical distances (Blitz et al. 1986; Fich et al. 1989), and by the HI disk-thickness method (Honma and Sofue 1997). Optical measurements of carbon stars have added many data points for outer rotation (Demers and Battinelli 2007). An accurate rotation velocity was obtained at a galactocentric radius of 13 kpc using VERA from parallax-proper motion measurements (Honma et al. 2007). These data were recently integrated to derive an accurate rotation curve of the Galaxy (Sofue et al. 2008). Here, we assume the galactocentric distance of the Sun to be 8.0 kpc and its circular rotation velocity of 200 km s$^{-1}$. 

Rotation curves are usually decomposed into several components, mostly into a bulge, disk and dark halo (Sofue and Rubin 2001). We have decomposed the newest rotation curve into a de Vaucouleurs bulge, exponential disk, and isothermal dark halo (Sofue et al. 2008). We showed that the outer rotation curve is reasonably fitted by an isothermal model. However, it was also pointed out that the halo model might not be unique: It is difficult to clearly discriminate one model from the other, e.g. isothermal model (Begeman et al. 1991) from NWF (Navarro et al. 1996) and Burkert models (Burkert 1995), because of the large scatter of data as well as for the limited radius within which the rotation curve is observed. Recent extensive observations of M31 showed that the rotation curve is similar to that of the Galaxy and can be decomposed into bulge, disk and an extended dark halo (Carignan et al. 2006).

Kinematics of satellites and member galaxies in the Local Group is a crew to estimate the dark halo, which is considered to be extended far outside the galactic disk and in the intracluster space. Kahn and Woltjer (1959) for the time argued for dark matter in the Local Group beyond the Galactic disk, obtaining a few $10^{12}\Msun$, by analyzing the motion of satellite galaxies.  There have been many works to use satellite galaxies to probe the mass and density distribution around galaxies (e.g., Einasto, et al. 1974; Ostriker et al. 1974; Zaritsky et al. 1989, 1997; Fich and Tremaine 1991;  Kochanek 1996; Kulessa and Lynden-Bell 1992;  Peebles et al. 2001; Prada et al. 2003). Recently, Li and White (2007), and van der Marel and Guhathakurta (2008) analyzed the radial velocities of satellite galaxies surrounding the Galaxy and M31, deriving a total mass of the Local Group of several $10^{12}\Msun$. Sawa and Fujimoto (2005) have computed probable past orbits of almost all major members of the Local Group under the condition that their positions and radial velocities are satisfied at the present time. They suggested a great encounter of the Galaxy and M31 group in the past. Cox and Loeb (2008) predict a future collision again of the Galaxy with M31.

In the present paper, we examine the behaviors of rotation curves calculated for different dark halo models in the outskirts of the Galaxy and in the Local Group. We compare the model rotation curves with a "pseudo rotation curve" of the Local Group, which combines the rotation curve of the Galaxy and radial velocities of the members of the Local Group.

\section{Galactic Rotation Curve}

\begin{figure} 
\bc
\includegraphics[width=8cm]{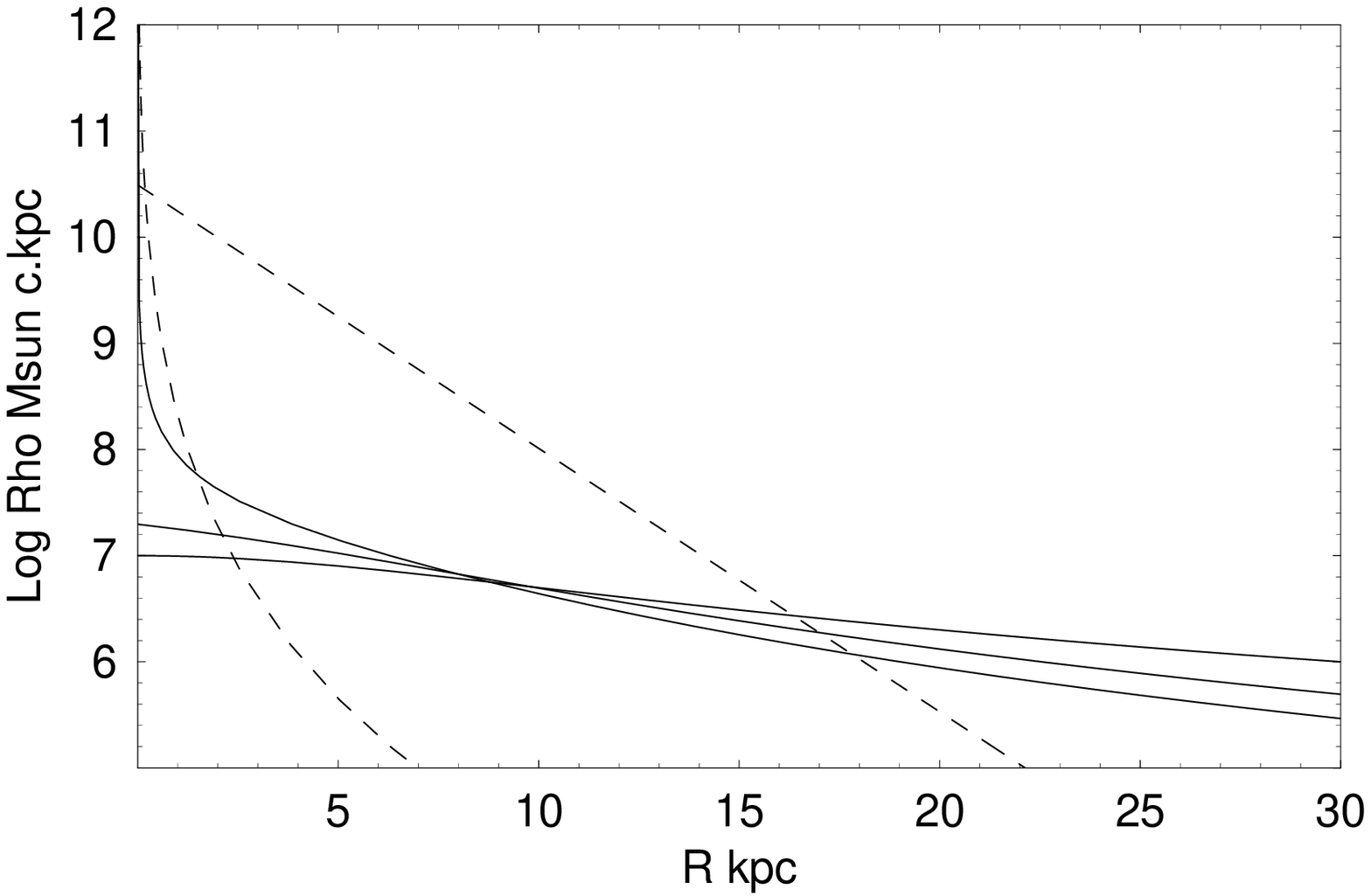} \\ 
\ec
\caption{Schematic plot of the volume density in the Galaxy for the approximate density profiles discussed in section 2. Inner and outer dashed lines are for the bulge and disk components, respectively. Solid lines from the top to bottom at $R=30$ kpc are contributions by the dark halos of isothermal, Burkert and NKF models, respectively. }
\label{fig-rho} 
\bc
\includegraphics[width=8cm]{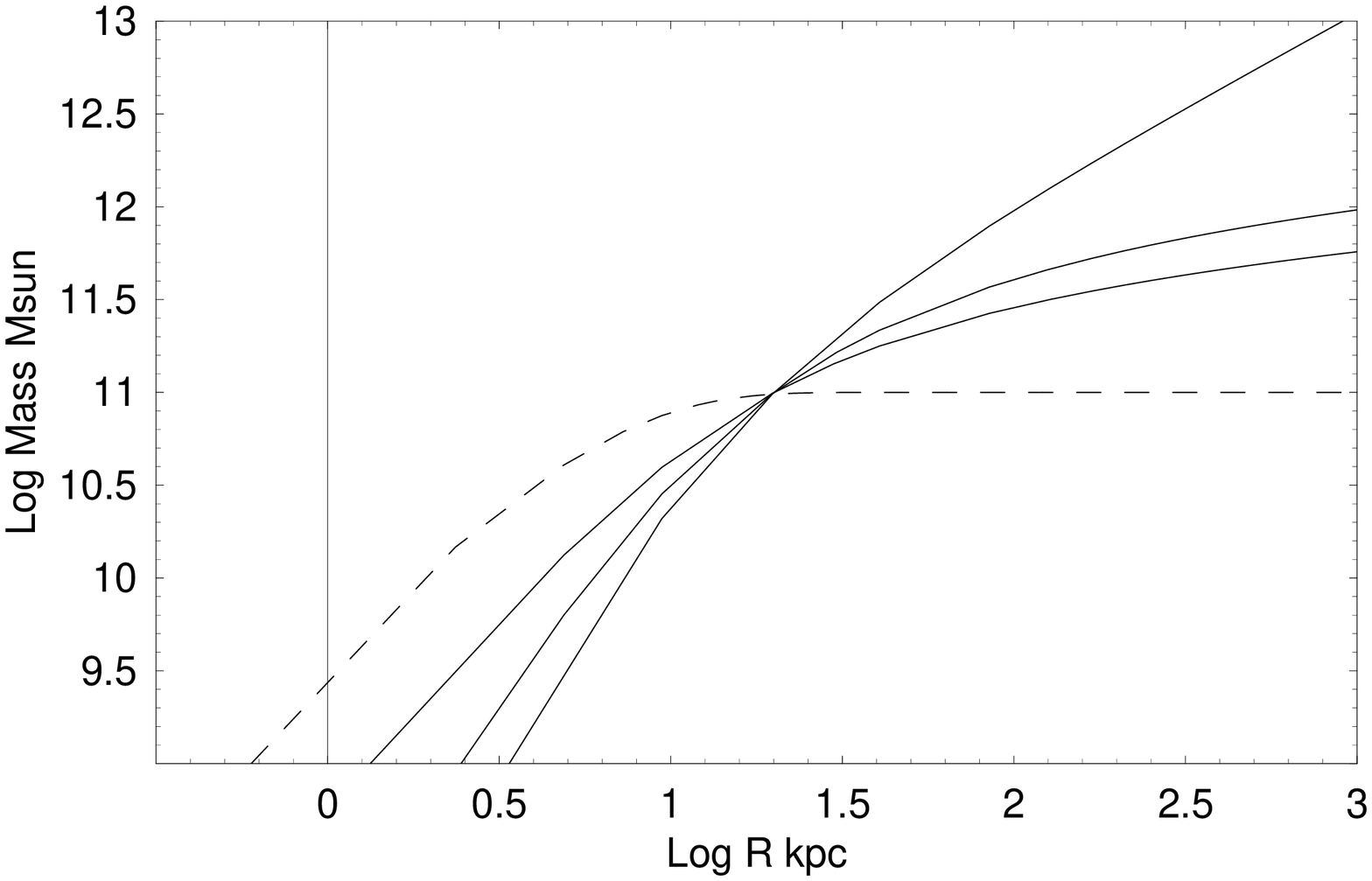} \\ 
\ec
\caption{Schematic plot of enclosed masses within radius $R$ for the dark halos (solid lines) and the disk (dashed line), corresponding to figure \ref{fig-rho}. }
\label{fig-mass} 
\bc
\includegraphics[width=8cm]{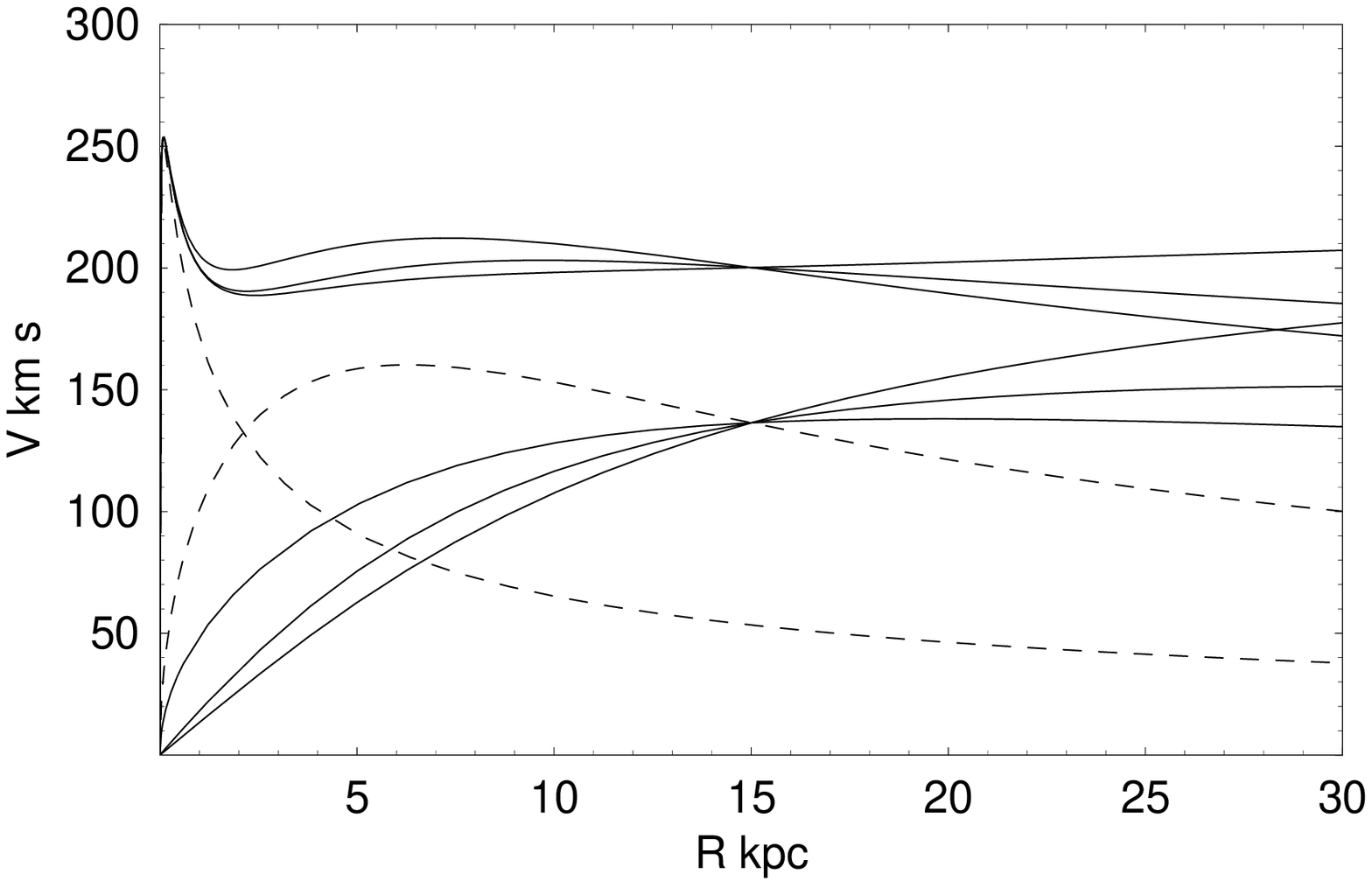} \\ 
\ec
\caption{Schematic rotation curves of the Galaxy for three dark halo models: isothermal, Burkert and NFW dark halo models, by solid lines from top to bottom at $R=30$ kpc. Lower curves show rotation velocities due to dark halos alone, and upper curves are for those including the bulge and disk. The inner and outer dashed lines are for the bulge and disk components, respectively. }
\label{fig-rcmodels} 
\end{figure} 

According to Sofue et al. (2008), we assume that the Galaxy is composed of three mass components: a bulge, disk and dark halo. The surface mass distribution of a galactic bulge is expressed by the de Vaucouleurs (1958) law, which is decomposed into the volume density distribution for calculating the circular rotational velocity. The surface mass distribution of the disk is given by a thin exponential disk, and the rotation velocity is calculated by using the modified Bessel functions (e.g. Freeman 1970). The volume density in the galactic disk is approximated by dividing the surface density by scale height, which increases with the radius, which is equal to 144 pc in the solar vicinity. In Sofue et al. (2008), we have shown  that the inner Galactic rotation curve is best-fitted by a combination of a bulge mass of $1.8\times 10^{10}\Msun$, its scale radius 0.5 kpc, disk mass $7\times10^{10}\Msun$ and scale radius 3.5 kpc, which are surrounded by a dark halo.

For the dark halo, we consider three cases of mass distributions: (1) semi-isothermal spherical distribution (Begeman et al. 1991), (2) NFW (Navarro, Frenk and White 1996), and (3) Burkert (1996) models. The density profiles are written, respectively, as
\be
\rho_{\rm iso} (R)={\rho_{\rm iso} ^0 \over  1+ (R/h)^2},
\label{eq_iso}
\ee 
\be 
\rho_{\rm NFW} (R)={\rho_{\rm NFW} ^0 \over (R/h)/[1+(R/h)^2]} ,
\label{eq-nfw}
\ee
and
\be \rho_{\rm Bur} (R)={\rho_{\rm Bur} ^0 \over [1+(R/h)][1+(R/h)^2)]} .
\label{eq-bur}
\ee 
Here, $h$ is the scale radius (core radius) of the dark halo. The circular rotation velocity is calculated by
\be
V_{\rm h} (R)=\sqrt{GM_{\rm h} (R)\over R},
\ee
where 
\be
M_{\rm h} (R)=4 \pi \int_0^R \rho_i(x) x^2 dx,
\ee
with $i=$iso, Bur, or NFW. 

A scale radius of $h=5.5$ kpc has been obtained for an isothermal halo, which best-fits the outer rotation curve after subtracting maximum bulge and disk contributions (Sofue et al. 2008). Scale radius on the order of 3.5 to 10 kpc have been used for NFW profiles by other authors (e.g. Hayashi and Chiba 2006; Seljak 2002). We adopt $h=10$ kpc commonly for the three models throughout this paper, as it is not a sensitive parameter for discussing the outer Galaxy.

Figure \ref{fig-rho} shows the density distributions for individual components of the bulge and disk, as well as those for the three different models of the dark halo. The parameters are the same as those adopted in Sofue et al. (2008) except for the halo scale radius ($h=10$ kpc). Figure \ref{fig-mass} shows the corresponding masses enclosed within a sphere of radius $R$. In these figures, the total masses of the disk and halo are adjusted so that they become equal to each other at a radius $R_{\rm c}$, which we call the disk-halo critical radius. We take $R_{\rm c}=15$ kpc in the present calculation. Obviously, the circular velocities corresponding to the disk and halo become approximately equal to each other at the critical radius. 

The circular velocity of the Galaxy is calculated by 
\be
V(R)^2=V_{\rm b} (R)^2+V_{\rm d} (R)^2+V_{\rm h} (R)^2,
\label{eq_rc}
\ee 
where $V_{\rm b} (R)$, $V_{\rm d} (R)$, and $V_{\rm h} (R)$ are the circular velocity corresponding to individual components of the bulge, disk, and dark halo, respectively.
Figure \ref{fig-rcmodels} shows the calculated rotation curves for the bulge, disk and halo as well as their sum, for the three isothermal, NFW and Burkert halo models.  The calculated rotation curve may be typical for a spiral galaxy having the standard structure, and mimics those obtained for M31 and other disk galaxies (Carignan et al. 2006; Hayashi and Chiba 2006). Rotation curves based on NFW density profiles have been obtained for various types of galaxies (Seljak 2002). 
 
\begin{figure*} 
\bc
\includegraphics[width=15cm]{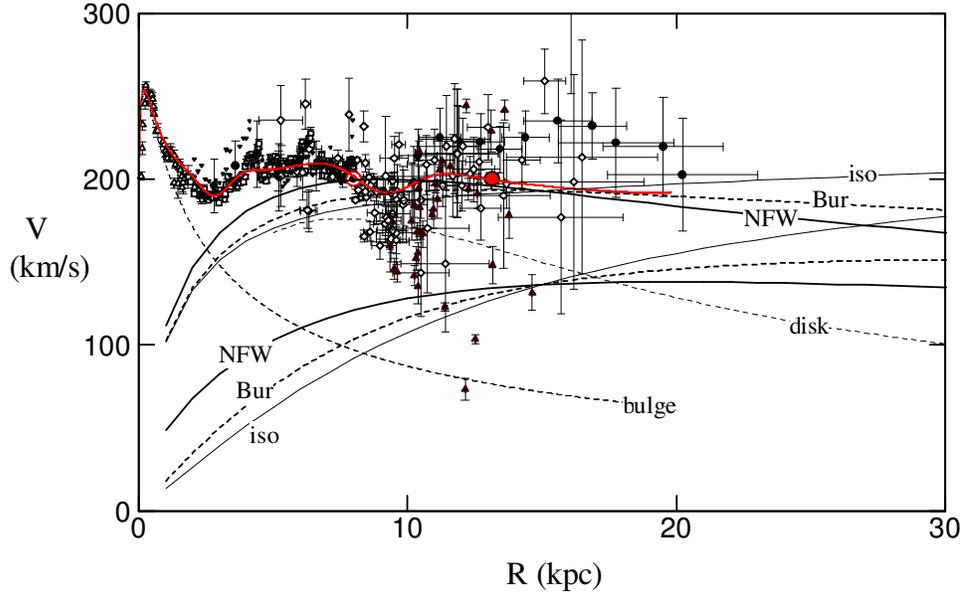} \\ 
\ec
\caption{Observed rotation curve of the Galaxy (Sofue et al. 2008). Rotation curves are shown for three dark halo models of scale radius $h=10$ kpc and critical radius $R_{\rm c}$=15 kpc. Thin solid line indicates the isothermal dark halo model, thick dashed lines Burkert, and thick solid lines the NFW model. For each model, the upper line indicates the sum of those for the dark halo and disk. The best-fit inner rotation curve is shown by grey line (Sofue et al. 2008). The bulge and disk contributions alone are indicated by thin dotted lines.}
\label{fig-rc30} 
\end{figure*}

Figure \ref{fig-rc30} shows the observed rotation curve of the Galaxy reproduced from Sofue et al. (2008), which is a compilation of observations in the decades from HI, CO, optical and VERA measurements. The observations are well fitted by a model rotation curve by the grey line (Sofue et al. 2008), which can be decomposed into three components: a de Vaucouleurs bulge of mass $M_{\rm b} =1.8 \times 10^{10}\Msun$ and scale radius $R_{\rm b} =0.5$ kpc, an exponential disk of mass $M_{\rm d} =7 \times 10^{10}\Msun$ and scale radius $R_{\rm d} =3.5 $ kpc, and an isothermal dark halo of scale radius $h=5.5$ kpc with the terminal flat rotation velocity of 200 \kms. 

In figure \ref{fig-rc30} we plot calculated rotation curves for the three dark halo models as shown in figure \ref{fig-rcmodels}. Thin solid line indicates isothermal dark halo model ({\it iso}), solid lines NFW model ({\it nfw}), and dashed lines Burkert model ({\it bur}). The curves are so adjusted that the velocity, or the enclosed mass, at $R_{\rm c}=15$ kpc is equal to each other, and the masses within $R_{\rm c}$ are taken to be equal to that of the exponential disk . For each model, the upper line indicates the sum of those for the dark halo and disk. The thin dashed lines are for the bulge and disk. 

In the inner Galaxy at $R<\sim 10$ kpc, the rotation velocity is predominantly determined by the bulge and disk contributions, and $R<0.5$ kpc it is almost determined by the bulge alone. The dark halo contribution is negligible in any models at these radii. Thus, it is hard to discriminate the models by comparing them with the observations in the inner Galaxy, although the data are most accurately obtained there.

Near the Galactic Center at $R<1$ kpc, the volume density of the NFW dark halo increases toward the nucleus inversely proportional to the radius (figure \ref{fig-rho}). However, the bulge density, obeying the de Vaucouleurs law, is kept sufficiently large compared to the dark halo density near the Galactic Center. The Burkert and isothermal halo models indicate a finite mild density plateau at the center, which is negligible compared to the disk and bulge densities. Accordingly, in any models, the contribution from the dark halo to the rotation velocity is negligible in the inner Galaxy. It is, therefore, not possible to distinguish the dark halo models by comparing with the inner rotation curve. 

On the other hand, in the outer Galaxy, the dark halo is the major mass component, and therefore, the rotation velocity is sensitive to the dark halo model. The outer rotation velocities for the three models are diverging as the radius increases, particularly at $R>\sim 20$ kpc. However, observational data are still restricted to the galactic disk at $R<20$ kpc, where the rotation curve is approximately flat. Also, the observations are largely scattered in the outer Galaxy, except for the recently added accurate value of $V=200\pm6$ \kms at $R=13.1\pm0.2$ kpc from VERA observations (Honma et al. 2007). Hence, it may still be too crude to use the galactic rotation curve in the present accuracy for discriminating the dark halo models, in so far as available observational kinematics using stars and interstellar gases are used.

\section{Pseudo Rotation Curve of the Local Group}

In order to discuss the dark halo and its mass distribution, kinematics beyond the Galactic disk is essential. There have been various works to use satellite and member galaxies in the Local Group for deriving the mass structure in the Local Group (the literature cited in Section 1).
We, here, examine a possibility to use radial velocities of the Local Group of galaxies as an extension of the rotation curve of the Galaxy. We use the observed radial velocities in a direct way for continuing the dynamical mass analysis of Galactic rotation curve. 

\begin{figure*}
\bc
\includegraphics[width=16cm]{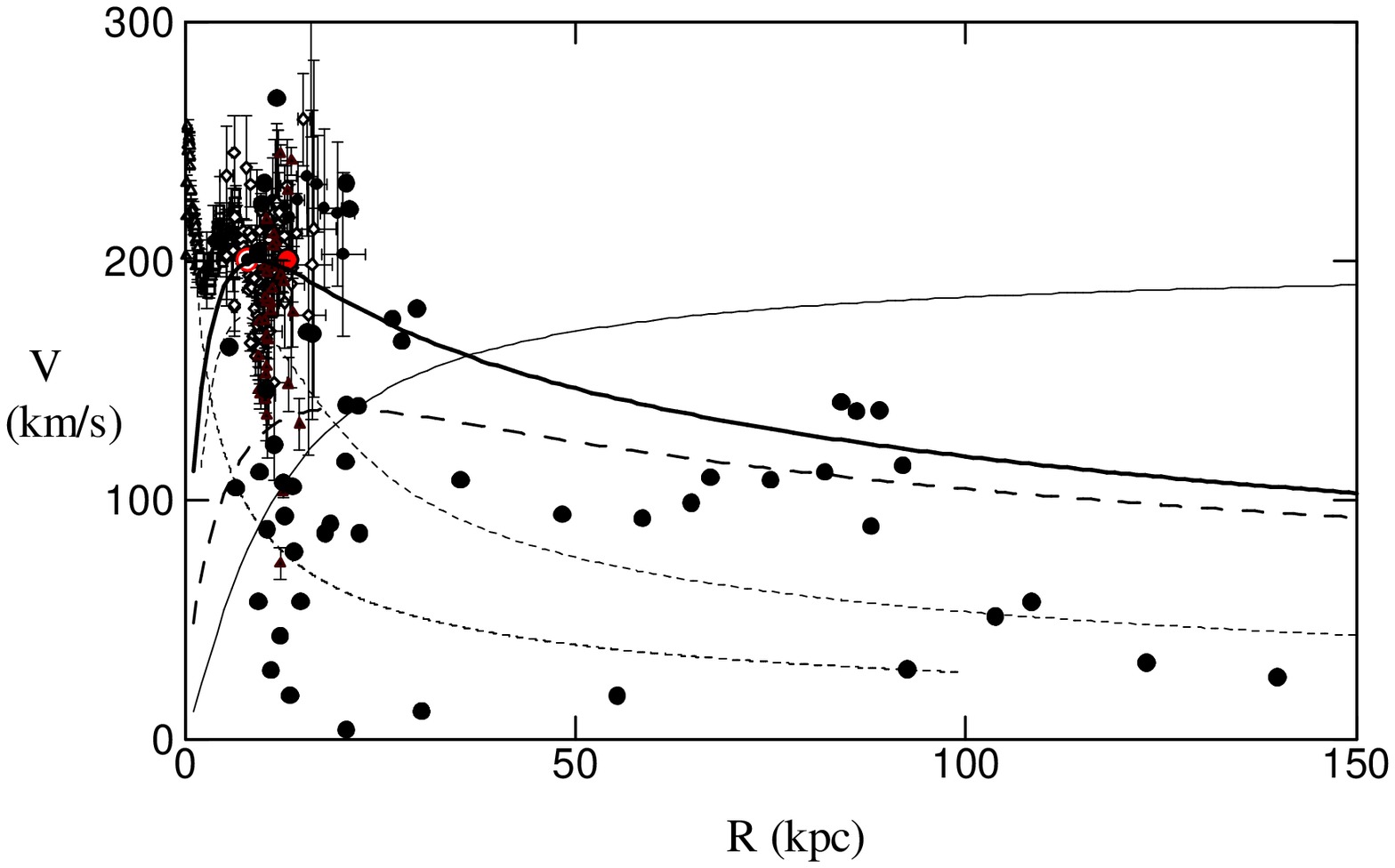} \\ 
\ec
\caption{The Galacto-Local Group rotation curve. The Galactic rotation curve at $R<20$ kpc are taken from Sofue et al. (2008) as in figure \ref{fig-rc30}. The data outside the Galactic disk are absolute values of Galacto-centric radial velocities taken from table 1 for globular clusters and satellite galaxies. The thick full line indicates the rotation curve for the NFW halo model (sum of the disk and halo) and dashed line is the NFW halo only. The asymptotic horizontal thin line represents the isothermal model. Note that the curve for the NFW model better traces the upper boundary of the plot than the isothermal model.}
\label{fig-150}  
\end{figure*}

\begin{figure*}
\bc
\includegraphics[width=16cm]{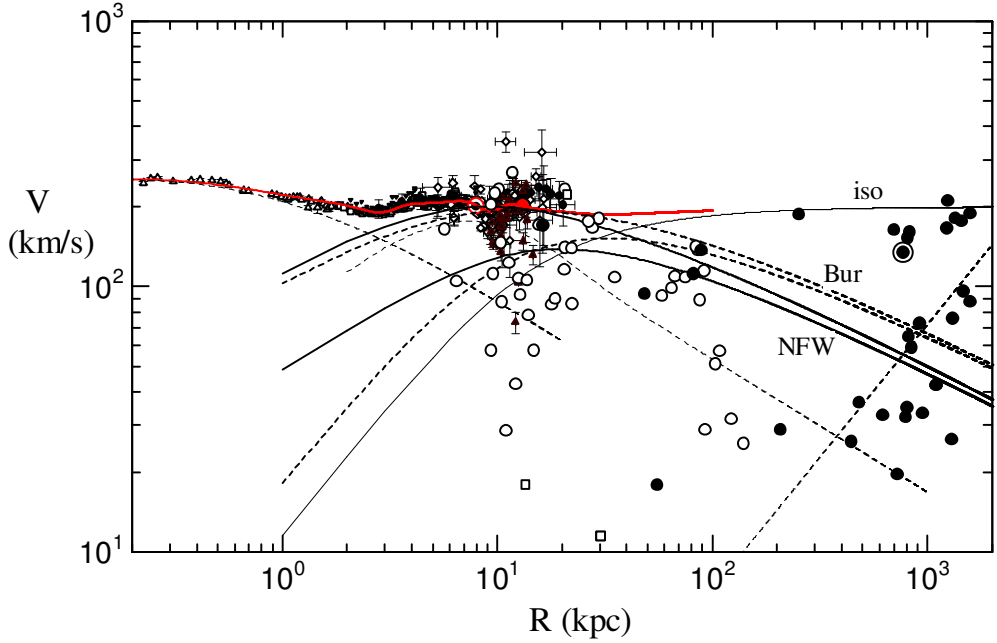} \\ 
\ec
\caption{Galacto-Local Group rotation curve. The data outside the Galactic disk are absolute values of Galacto-centric radial velocities taken from table. Open circles and squares are globular clusters and satellites of the Galaxy, and filled circles are Local Group members. The filled circle doubled by a big open circle at 770 kpc indicates M31.  Model rotation curves calculated for three different halo models, same as in figure \ref{fig-rc30}, are shown together. Solid and dashed lines are for NFW and Burkert models, respectively. The thin asymptotically horizontal line represents the isothermal model for terminal velocity 200 \kms, respectively. Hubble flow for $H_0=72$ \kms Mpc$^{-1}$ is shown by a dashed line near the right-bottom corner.}
\label{fig-vgc}   
\end{figure*}

In figure \ref{fig-150} we plot the absolute values of the galacto-centric radial velocities $V_r=|V_{\rm GC}|$ of outer globular clusters and members of the Local Group against their galacto-centric distances $R_{\rm GC}$ for objects within a distance of 150 kpc. Calculated rotation curves for the NFW model is shown by the thick line, and individual components are shown by dashed lines. The isothermal halo model is shown by the upper thin line, which is horizontal at large radii. The upper boundaries of the plot is well fitted by the rotation curve for the NFW model. In figure \ref{fig-vgc} we plot the same for entire region of the Local Group up to 1 Mpc.   For each object, the galacto-centric distance $R_{\rm GC}$ was calculated from galactic coordinates and helio-centric distance for $R_0=8$ kpc, and $V_r$ was calculated from observed helio-centric radial velocity by correcting for the solar motion of $V_0=200$ \kms after transforming to LSR velocity. We call these diagrams the "pseudo rotation curve" of the Local Group, or "Galacto-LG (GLG) rotation curve". These may be compared to similar plots of velocity differences against spatial separations of satellite galaxies for many pair galaxies as obtained by Zaritsky et al. (1997) and Prada et al. (2003).   

Used galaxies and globular clusters with their longitude, latitude, distances and velocities are listed in table 1. The velocity and position data were adopted from  Kulessa and Lynden-Bell (1992), Sawa and Fujimoto (2005) which is mainly based on those of Mateo (1998), Kochanek et al. (1996), and Kopylov et al. (2008). 

\begin{table*}
\bc
\caption{Globular clusters and satellite galaxies} 
\begin{tabular}{llllllll}
\hline\hline 
\\
Name & $l $ & $b$ & $R_{\rm helio}$ & $v_{\rm helio}$ & $v_{\rm LSR}$ & $R_{\rm GC}$ &$V_{\rm GC}$ \\ 
  & $(^\circ)$ & $(^\circ)$ & (kpc) & (\kms)& (\kms) & (kpc) &(\kms) \\
 \\
\hline 
\\
---KL sample---\\
\\
 NGC   288 &  151.0 &  -89.0 &   12.3 &  -48.2 &  -59.0 &   14.8 &  -57.3 \\
 NGC   364 &  301.0 &  -46.0 &   10.2 &  217.0 &  206.5 &   10.5 &   87.4 \\
 NGC  1261 &  270.0 &  -52.0 &   16.1 &   51.0 &   37.5 &   18.0 &  -85.6 \\
 NGC  1851 &  244.0 &  -35.0 &   16.4 &  306.0 &  286.6 &   20.7 &  139.4 \\
 NGC  1904 &  277.0 &  -29.0 &   20.0 &  193.0 &  177.3 &   20.7 &    3.7 \\
 NGC  2298 &  245.0 &  -16.0 &   17.9 &  108.0 &   88.4 &   22.4 &  -85.8 \\
 NGC  2419 &  -26.0 &  180.0 &   84.7 &   20.0 &   26.6 &   92.0 &  114.3 \\
 NGC  1808 &  282.0 &  -11.0 &   11.5 &   98.0 &   84.9 &   12.6 & -107.1 \\
 NGC  3201 &  277.0 &    8.0 &    9.7 &  477.0 &  464.4 &   11.8 &  267.8 \\
 NGC  4147 &  252.0 &   77.0 &   20.2 &  177.0 &  181.7 &   22.2 &  138.9 \\
 NGC  4590 &  299.0 &   35.0 &   10.2 &  -84.0 &  -88.8 &   10.2 & -232.1 \\
 NGC  5024 &  333.0 &   80.0 &   18.1 &  -80.0 &  -74.3 &   18.6 &  -90.0 \\
 NGC  5272 &   42.0 &   79.0 &   12.4 & -141.0 & -131.0 &   13.8 & -105.5 \\
 NGC  5466 &   42.0 &   74.0 &   15.3 &  119.9 &  133.2 &   15.7 &  170.1 \\
 NGC  5634 &  342.0 &   49.0 &   17.5 &  -41.9 &  -37.4 &   14.0 &  -78.0 \\
 NGC  5694 &  331.0 &   30.0 &   26.1 & -152.0 & -148.3 &   20.7 & -232.2 \\
 NGC  5824 &  333.0 &   22.0 &   17.1 &  -38.0 &  -38.5 &   11.4 & -122.7 \\
 NGC  6341 &   68.0 &   45.0 &   10.0 & -120.5 & -102.6 &   11.0 &   28.6 \\
 NGC  6715 &    6.0 &  -14.0 &   13.0 &  133.0 &  143.4 &    5.7 &  163.7 \\
 NGC  6779 &   63.0 &    8.0 &    9.7 & -138.1 & -119.1 &    9.4 &   57.4 \\
 NGC  6864 &   20.0 &  -26.0 &   11.7 & -175.0 & -166.2 &    6.5 & -104.7 \\
 NGC  6934 &   52.0 &  -19.0 &   12.0 & -387.0 & -372.5 &    9.8 & -223.5 \\
 NGC  6981 &   35.0 &  -33.0 &   12.9 & -309.0 & -299.5 &    9.4 & -203.3 \\
 NGC  7006 &   64.0 &  -19.0 &   32.1 & -364.0 & -349.6 &   29.7 & -179.6 \\
 NGC  7078 &   65.0 &  -27.0 &   10.3 &  -10.0 &  -15.6 &   10.4 &  145.9 \\
 NGC  7089 &   53.0 &  -36.0 &   10.5 &   -5.0 &  -17.9 &    9.6 &  111.3 \\
 NGC  7492 &   53.0 &  -63.0 &   21.3 & -188.5 & -188.2 &   20.6 & -115.7 \\
 Pal     1 &  130.0 &   19.0 &   22.2 &    3.0 &   21.3 &   27.8 &  166.1 \\
 Pal     2 &  170.0 &   -9.0 &   27.5 & -133.0 & -142.5 &   35.3 & -108.2 \\
 Pal     3 &  240.0 &   42.0 &  100.6 &   84.0 &   77.7 &  103.8 &  -51.0 \\
 Pal     4 &  202.0 &   72.0 &  105.9 &   75.0 &   80.2 &  108.5 &   57.1 \\
 Pal     5 &    1.0 &   45.0 &   16.5 &  -55.4 &  -45.3 &   12.2 &  -42.8 \\
 Pal    12 &   31.0 &  -48.0 &   15.6 &    9.0 &   23.9 &   12.8 &   92.9 \\ 
 Pal    13 &   87.0 &  -42.0 &   25.7 &   13.0 &   27.2 &   26.6 &  175.6 \\
 Pal    14 &   29.0 &   42.0 &   63.5 &    0.0 &   19.9 &   58.6 &   92.0 \\
 Pal    15 &   19.0 &   24.0 &   74.1 &   32.0 &   49.7 &   67.3 &  109.2 \\
AM -1 &  258.0 &  -48.0 &  121.8 &  116.0 &   99.2 &  123.2 &  -31.7 \\
Erid. &  218.0 &  -41.0 &   81.0 &  -25.0 &  -44.0 &   86.0 & -136.9 \\
Draco &   86.0 &   35.0 &   75.0 & -289.0 & -271.5 &   75.0 & -108.1 \\
Ursa Min. &  105.0 &   45.0 &   63.0 & -249.0 & -235.2 &   64.9 &  -98.6 \\
Sculptor&  288.0 &  -83.0 &   84.0 & -107.0 & -117.5 &   84.1 & -140.7 \\
Fornax &  237.0 &  -66.0 &  138.0 &   55.0 &   42.6 &  140.0 &  -25.6 \\
Carina &  260.0 &  -22.0 &   91.0 &  230.0 &  211.5 &   92.6 &   28.9 \\

\\
---K sample---\\ 
\\
Rup 106 &  301.0 &   12.0 &   24.0 &  -44.0 &  -53.7 &   21.1 & -221.4 \\
Pal   8 &   14.0 &   -7.0 &   21.0 &  -38.0 &  -30.0 &   13.5 &   18.0 \\ 
NGC  6229 &   74.0 &   40.0 &   31.0 & -154.0 & -135.7 &   30.3 &   11.5 \\ 

\hline 
\end{tabular}
\ec

\label{tab1}
\end{table*} 

\begin{table*}
\bc 
\begin{tabular}{llllllll}
---(Table 1 continued)---\\
\hline\hline 
\\
Name & $l $ & $b$ & $R_{\rm helio}$ & $v_{\rm helio}$ & $v_{\rm LSR}$ & $R_{\rm GC}$ &$V_{\rm GC}$ \\ 
  & $(^\circ)$ & $(^\circ)$ & (kpc) & (\kms)& (\kms) & (kpc) &(\kms) \\
 \\
\hline 
\\
---SF sample---\\
\\
WLM DDO 221 & 75.9 &  -73.6 &  925.0 & -123.0 & -127.1 &  924.5 &  -72.4 \\
NGC 55 &  332.7 &  -75.7 & 1480.0 &  124.0 &  118.4 & 1478.3 &   95.7 \\
IC 10 UGC192&  119.0 &   -3.3 &  825.0 & -342.0 & -334.3 &  828.9 & -159.7 \\
NGC 147 DD3 &  119.8 &  -14.3 &  725.0 & -193.0 & -187.8 &  728.9 &  -19.6 \\
NGC 185 UGC3966 &120.8 & -14.5 &  620.0 & -204.0 & -199.0 &  624.0 &  -32.7 \\
NGC 205 M110 & 120.7 &  -21.1 &  815.0 & -229.0 & -225.1 &  818.8 &  -64.6 \\
M32 NGC 221&  121.2 &  -22.0 &  805.0 & -197.0 & -193.5 &  808.9 &  -34.9 \\
M31 NGC 224&  121.2 &  -21.6 &  770.0 & -297.0 & -293.1 &  773.9 & -134.0 \\
M33 NGC 598&  133.6 &  -31.3 &  840.0 & -181.0 & -182.5 &  844.7 &  -58.8 \\
SMC NGC 292&  302.8 &  -44.3 &   58.0 &  148.0 &  138.2 &   55.4 &   17.9 \\ 
LGS3 Pisces &  126.8 &  -40.9 &  810.0 & -272.0 & -273.0 &  813.7 & -152.0 \\
NGC 1613 DDO8&  129.8 &  -60.6 &  700.0 & -234.0 & -238.9 &  702.6 & -163.5 \\
Phoenix &  272.2 &  -68.9 &  445.0 &   56.0 &   46.0 &  445.0 &  -26.0 \\ 
EGB0427+63 UGCA92&144.7 &10.5 & 1300.0 &  -87.0 &  -87.1 & 1306.4 &   26.5 \\
LMC &  280.5 &  -32.9 &   49.0 &  274.0 &  258.9 &   48.4 &   93.8 \\ 
Leo A DDO69 &  196.9 &   52.4 &  690.0 &   26.0 &   30.8 &  694.7 &   -4.7 \\
Sextans B DDO70&233.2 &   43.8 & 1345.0 &  303.0 &  295.7 & 1348.5 &  180.2 \\
NGC 3109 &  262.1 &   23.1 & 1250.0 &  404.0 &  392.1 & 1251.0 &  209.9 \\
Antlia &  263.1 &   22.3 & 1235.0 &  361.0 &  349.1 & 1235.9 &  165.4 \\
Leo I &  226.0 &   49.1 &  250.0 &  286.0 &  280.7 &  253.7 &  186.5 \\
Sextans A DDO75& 246.2 & 39.9 & 1440.0 &  325.0 &  316.6 & 1442.5 &  176.2 \\
Sextans &  243.5 &   42.3 &   86.0 &  277.0 &  269.4 &   89.0 &  137.0 \\
Leo II DDO93 &  220.2 & 67.2 &  205.0 &   76.0 &   78.9 &  207.5 &   28.9 \\
GR8 DDO155 &  310.7 &   77.0 & 1590.0 &  215.0 &  222.2 & 1588.8 &  188.1 \\ 
Draco DDO208 &   86.4 &   34.7 &   82.0 & -293.0 & -275.6 &   82.0 & -111.5 \\
Sagittarius  &    5.6 &  -14.1 &   24.0 &  140.0 &  150.3 &   16.4 &  169.2 \\
Sgr DIG UKS1927-177& 21.1 & -16.3 & 1060.0 & -79.0 & -68.8 & 1052.8 &   0.3 \\
NGC 6822 DDO209&   25.3 &  -18.4 &  490.0 &  -54.0 &  -44.5 &  483.2 &  36.6 \\
DDO210 Aquarius& 34.0 &  -31.3 &  800.0 & -137.0 & -127.8 &  794.4 &  -32.2 \\
IC 5152 &  343.9 &  -50.2 & 1590.0 &  124.0 &  123.0 & 1585.1 &   87.5 \\
UKS2323-326 UGCA438& 11.9 & -70.9 & 1320.0 & 62.0 &   62.1 & 1317.5 &   75.6 \\
Pegasus DDO216&   94.8 &  -43.5 &  955.0 & -182.0 & -177.9 &  955.5 &  -33.3 \\
UGC 4879 VV124&  164.7 &   42.9 & 1100.0 &  -80.0 &  -81.3 & 1105.7 &  -42.5 \\

\hline 
\end{tabular}
\ec 

KL sample: Kulessa and Lynden-Bell (1992). K sample: Kochanek (1996): Globular clusters and satellite galaxies. SF sample: Local Group galaxies from Sawa and Fujimoto (2006), whose list is mainly based on Mateo (1998).  See the literature for original references. VV124: Kopylov et al. (2008). 

\label{tab1-con}
\end{table*}

In figures and \ref{fig-150} and \ref{fig-vgc} we show the calculated rotation curves for the three dark halo models: isothermal, NFW and Burkert profiles. The dark halo models are so chosen that they are smoothly connected to the inner composite rotation curve for the best-fit bulge and disk models, and that the flat part at $R\sim 10-20$ kpc is approximately reproduced. The scale radius of the dark halo is taken to be $h=10$ kpc. The total mass of the dark halo within the critical radius $R_{\rm c}=15$ kpc in each model is taken to be equal to the disk mass in the same radius. 

As equations (\ref{eq-nfw}) and (\ref{eq-bur}) indicate, the NFW and Burkert profiles are similar to each other at radii sufficiently greater than the scale radius $h$, e.g. at $R\gg h=10$ kpc. Accordingly, the rotation curves corresponding to these two models are similar at $R>30$ kpc, except for a small difference arising from the difference in the profiles, and therefore the total masses, at  $R< \sim h$.

If we look into the local area around the Galaxy at $R<\sim 150$ kpc as in figure \ref{fig-150}, the NFW well traces the upper envelope of the  pseudo rotation curve. 
On the other hand, if look into the plot at larger scales as in figure \ref{fig-vgc},  the isothermal model can approximately trace the upper boundary of the entire observations in the Local Group. Interestingly, M31, one of the major two massive galaxies in the Local Group, lies on this curve. 

These facts imply that the mass distribution in the Galaxy is an isolated system from the other members in so far as it is observed within $\sim 150$ kpc radius. Hence, if it is gravitationally bound to the Local Group, there must exist a much larger amount of dark mass than the dark halo of the Galaxy, filling the entire Local Group. 

\section{Discussion}

\subsection{Inner vs Outer Rotation Curves}

We have compared rotation curves based on the three dark halo models, isothermal, NFW, and Burkert, with the observed Galacto-Local Group rotation curve. In the inner Galaxy at $R<\sim 10$ kpc, the disk and bulge are the major constituents of the mass, and therefore, the rotation velocity is little affected by the dark halo. In the outer disk, the dark halo contribution becomes significant. However, the shape of the rotation curve is not sensitive to the halo models at $R\sim 10-20 $ kpc, which is almost flat in any model, and the discrimination among the models is again difficult. 

We here note that, for such reasons, there have been many studies utilizing dwarf and low-surface brightness galaxies, which are thought to be dark matter dominant (e.g. van den Bosch et al. 2000; Gentil et al. 2007; Jiminez et al. 2003). However, the major dynamical constituents of the Local Group are M31 and our Galaxy, two orders of magnitude more massive than dwarfs, where the bulge and disk are dominant within the optical radii at $<R_{\rm c} \sim 15$ kpc. It is, therefore, essential to investigate outer halo kinematics in order to discuss the dark halo in such a big galaxy as the Milky Way. For this purpose, we constructed a Galacto-Local Group rotation curve as shown in figures \ref{fig-vgc} and \ref{fig-150}. 

\subsection{NFW/Burkert model and the Mass of the Galaxy}

The upper boundary of the pseudo rotation curve within $\sim150$ kpc radius is well fitted by the NFW and Burkert models, as shown in figure \ref{fig-150}. In these models, the total mass of the Galaxy involved within a radius 770 kpc, the distance to M31, is $8.7\times 10^{11}\Msun$, and the mass within 385 kpc is $4.4 \times 10^{11}\Msun$, respectively. These values may be adjustable within a range less than $10^{12}\Msun$ by tuning the parameters $h$ and $R_{\rm c}$, so that the flat galactic rotation at $R\sim 10-20$ kpc can be approximated. 

In either case, the Galaxy's mass is by an order of magnitude smaller than the total mass required to gravitationally bind the Local Group. If we assume that the two galaxies are gravitationally bound to the Local Group, the total mass is estimated to be on the order of $M_{\rm tot} \sim V^2 R/G \sim 4.8 \times 10^{12}\Msun$. Here, the mutual velocity of the two galaxies is taken as $V \sim \sqrt{3} V_{\rm GC}$ with $V_{\rm GC}=-134$ \kms for M31, and the radius of the orbit is $R=385$ kpc, where the factor $\sqrt{3}$ was adopted as a correction for the freedom of motion. It is required that the Local Group contains a dark matter core of mass comparable to this binding mass, $\sim 4.8 \times 10^{12}\Msun$. The mean mass density required to stabilize the Local Group is, thus, estimated to be 
$\rho_{\rm LG} \sim 2 \times 10^{-5}\Msun{\rm pc}^{-3}$.   

We here define the boundary of the Galaxy as the radius, at which the density of the galactic dark halo becomes equal to the density of dark matter of the Local Group. The thus defined radius is $R_{\rm G}\sim 100$ kpc for the NFW model. The Galaxy's mass within this boundary is $M_{\rm G: NFW}= 3\meleven$ (figure \ref{fig-mass}).


The high-velocity ends of the pseudo rotation curve for the entire Local Group up to $\sim 1.5$ Mpc in figure \ref{fig-vgc} are well represented by an isothermal dark halo model. The estimated total mass $M \sim 4.8 \times 10^{12}\Msun$  for the terminal flat velocity of $V=134 \sqrt{3} \sim 230$ \kms may be compared with the mass of the Local Group derived by Li and White (2007) of $5.3\times 10^{12}\Msun$, and that by van der Marel and Guhathakurta (2008) of $5.6\times10^{12}\Msun$.
 
The mean dark matter density of the Universe is on the order of $\rho_0\sim (H_0 R)^2 R/G /(4\pi R^3/3)\sim 2\times 10^{-29}{\rm g ~cm^{-3}} \sim 1.2\times10^{-6}\Msun {\rm pc}^{-3} $ for $H_0=72$ \kms kpc$^{-1}$. Total mass of this "uniform" dark matter is $\sim 1.2 \meleven$ inside a sphere of radius 385 kpc, and $\sim 1 \mtwelve$ in 770 kpc, which is small enough compared to the dynamical mass of the Local Group.

The ratio of barionic mass (luminous mass) of a galaxy to the dark matter mass gives important information for the formation scenario of galaxies. The upper limit to the barionic mass of the Galaxy is approximately represented by the disk plus bulge mass, which was estimated to be $M_{\rm barion}\simeq 0.83\times 10^{11}\Msun$ by fitting of the inner rotation curve using de Vaucouleurs bulge and exponential disk (Sofue et al. 2009). This yields an upper limit to the barionic-to-dark matter mass ratio of the Galaxy within radius $R$ as $\Gamma(R)=M_{\rm barion}/\left\{M_{\rm total} (R)-M_{\rm barion}\right\}\simeq 0.38$ for $R=100$ kpc, and $\Gamma\simeq 0.23$ for $R=385$ kpc. These values are on the same order of the cosmological value of 0.2 from WMAP (Spergel et al. 2003). We may consider that the barionic mass of galaxies in the Local Group is represented by those of the Galaxy and M31, which is about twice the Galaxy's barionic mass. Then, the upper limit to the barion-to-dark matter mass ratio in the Local Group is estimated to be $\Gamma \sim 0.036$, which is only 0.2 times the cosmic value. It is expected that approximately 80\% of barionic matter of the Local Group exists in the intergalactic space without being captured to the present galaxies.


\subsection{The Galaxy borders M31 via Dark Halos}

 HI observations by Carignan et al. (2006) showed that M31's rotation curve is flat till a radius 35 kpc. They estimate the  mass of M31 within 35 kpc to be $3.4\meleven$. This is comparable to the total mass of the Milky Way Galaxy within 35 kpc of $3.2\meleven$. We may consider that the Galaxy and M31 have similar mass profiles.

\begin{figure} 
\bc
\includegraphics[width=8cm]{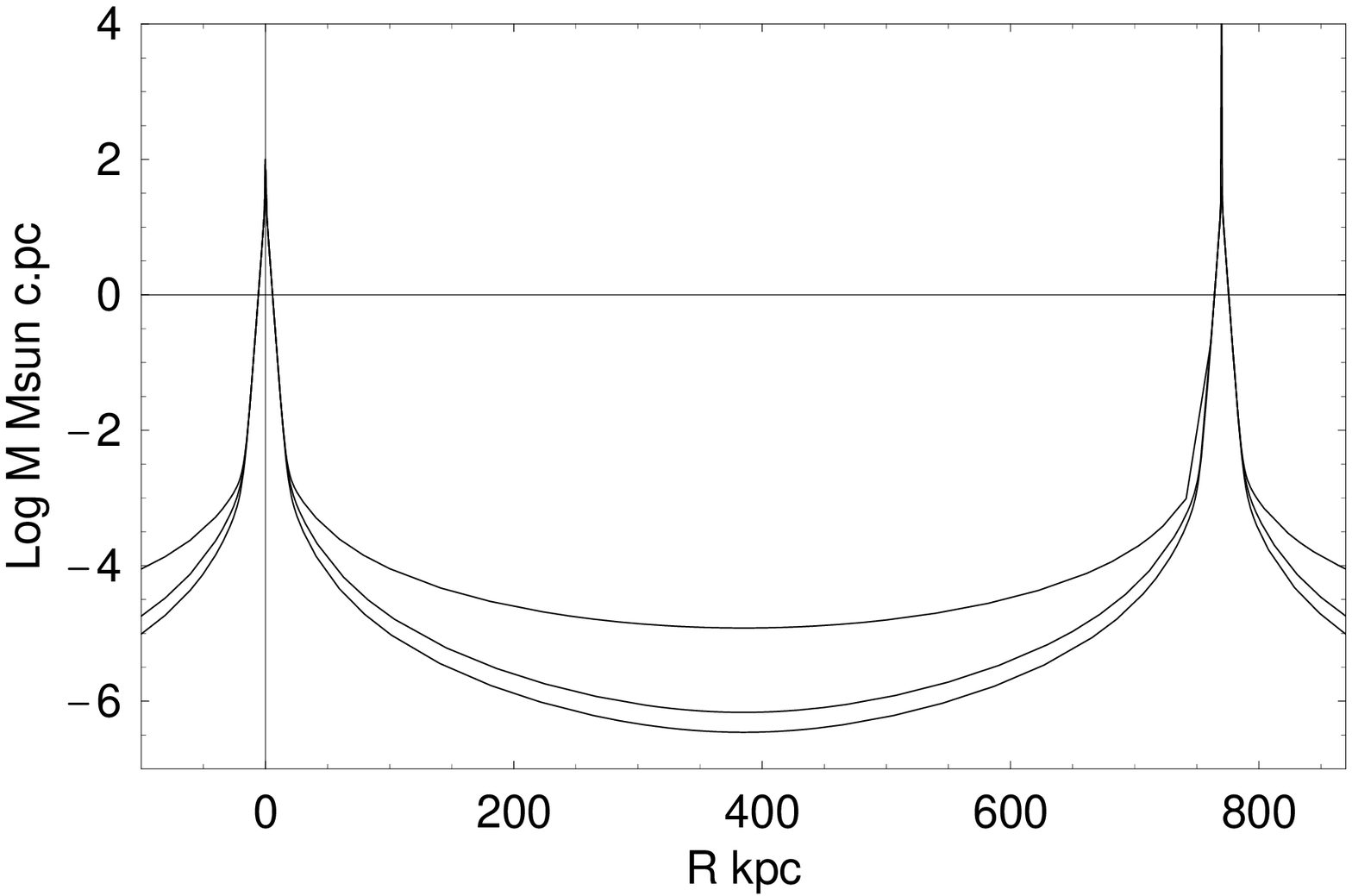} \\ 
\includegraphics[width=8cm]{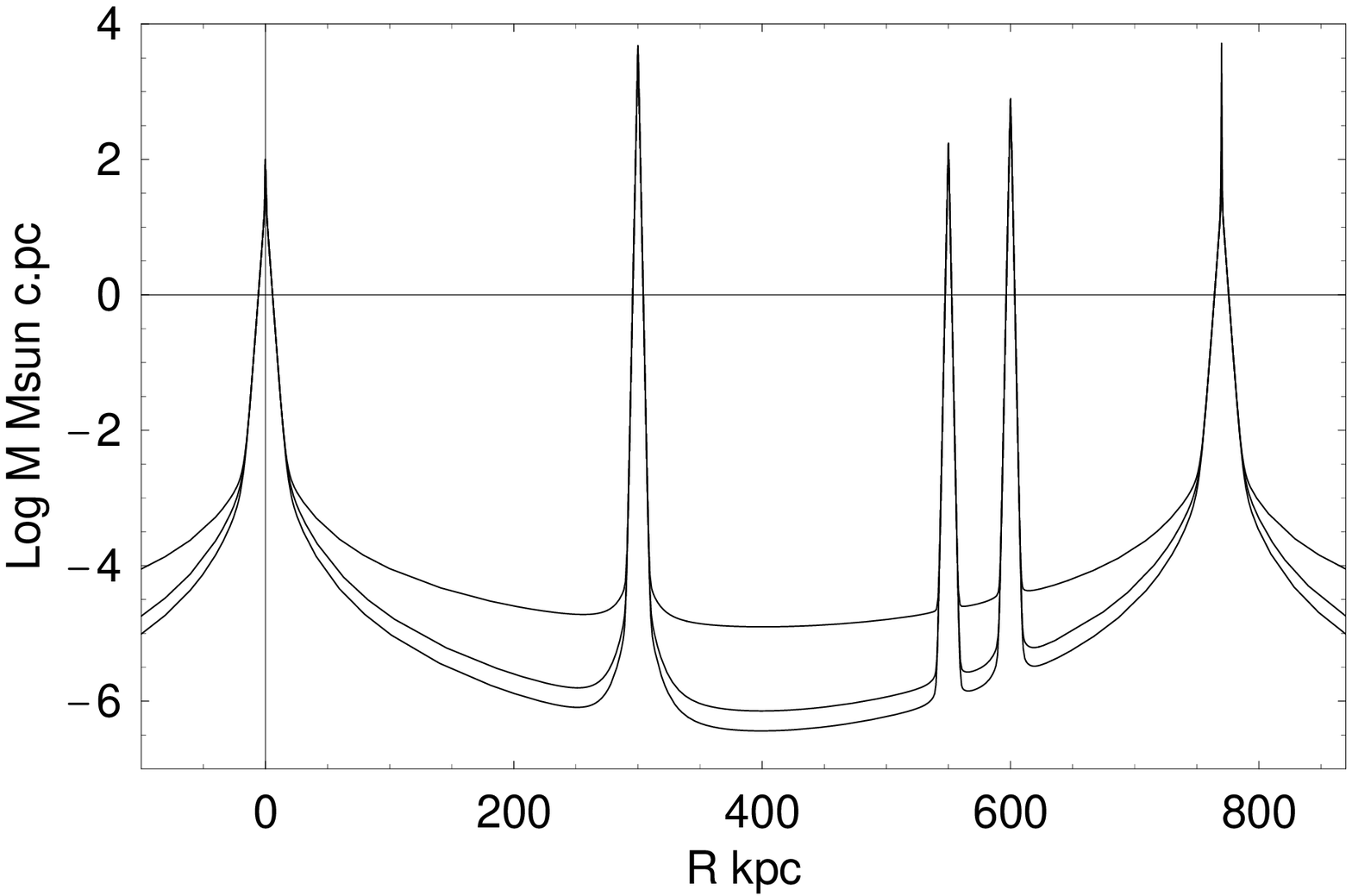} \\  
\ec
\caption{Top panel: Density profiles between the Milky Way and M31 for different dark halo models - isothermal, Burkert and NFW from top to bottom. Each galaxy is assumed to have the same density profile as in figure \ref{fig-rho}. The isothermal dark halo gravitationally binds the two galaxies. 
Bottom: Three small galaxies of a half scale radii and  masses one to two orders of magnitudes less are located on the line connecting the two giant galaxies. 
}
\label{fig-mwm31} 
\end{figure}

Figure \ref{fig-mwm31} shows calculated density profiles along a line crossing the centers of the Galaxy and M31, where the two galaxies are separated by 770 kpc and are assumed to have the same density profiles as that in figure \ref{fig-rho}. Considering that the Galaxy and M31 are embedded in dark halos, both galaxies border by the dark matter, and are dynamically bound by the Local Group dark mass. 

\begin{figure} 
\bc 
\includegraphics[width=8cm]{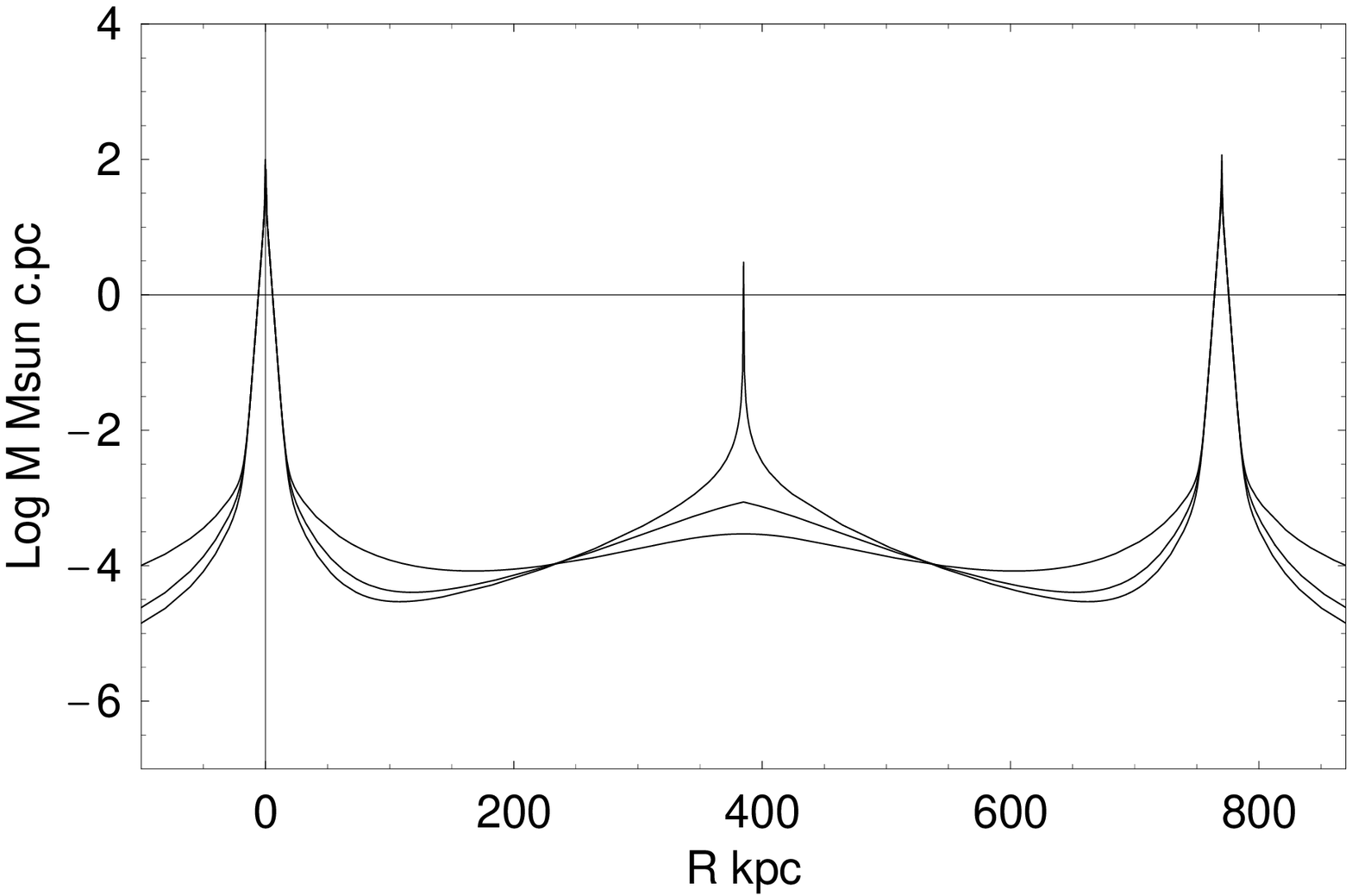} \\ 
\includegraphics[width=7cm]{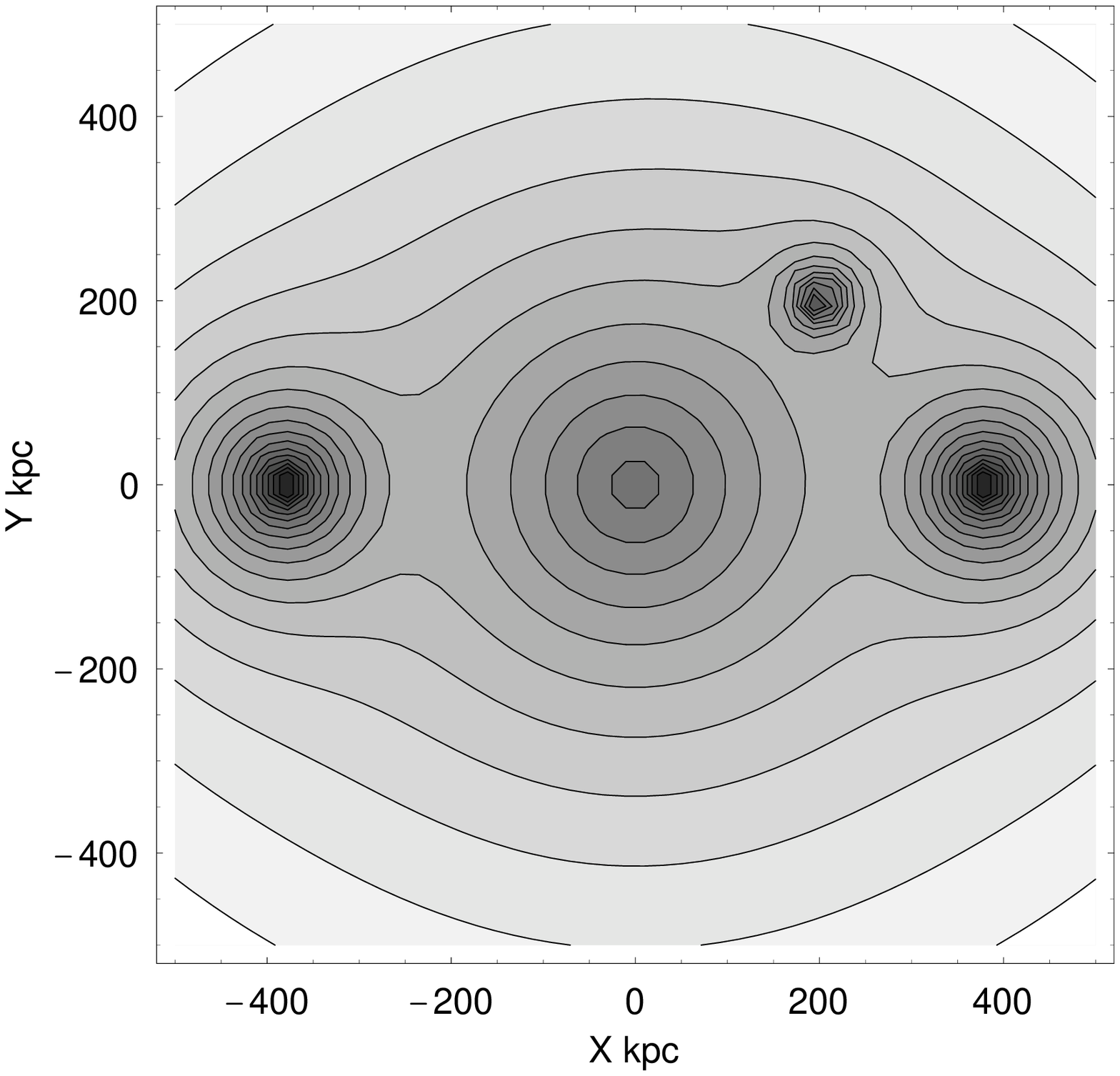} \\ 
\ec
\caption{ Top panel: Same as figure \ref{fig-mwm31}, but the Galaxy and M31 are located on both sides of a dark matter core of the Local Group whose core radius is $h=100$ kpc, and integrated mass within a sphere of 385 kpc is adjusted to  $5 \mtwelve$, so that the dark matter can bind the Local Group.
Bottom panel: Schematic illustration of dark halos of the Galaxy, M31, and an arbitrary dwarf galaxy embedded in the Local Group dark core.
}
\label{fig-darklg} 
\end{figure} 

If the halo profile is either Burkert or NFW, the galaxies are not massive enough to bind each other, and cannot stabilize the Local Group.   An extended massive component of intracluster dark matter of several$\mtwelve$ is required. The bottom panel in figure \ref{fig-mwm31} illustrates a density profile for such a case that there exists a dark matter core  at the midpoint between the Galaxy and M31. For each model, the core radius is  $h=100$ kpc and the integrated mass within a sphere of radius 385 kpc is adjusted to $5 \mtwelve$, a sufficient mass to bind the Local Group.

In either model, less massive galaxies are alike floating islands in the dark matter ocean between the two continents (M31 and the Galaxy). In figure \ref{fig-mwm31} we illustrate a possible case where three small galaxies are embedded in dark halos of two big galaxies. The small galaxies are given one or two orders of magnitudes less masses, and are located at random on the line connecting the big galaxies. 

It is known that rotation curves of dwarfs and latest type galaxies increase monotonically toward their edges, as is typically observed for M33 (Corbelli and Salucci 2000). We may speculate that M33 borders M31 by their dark halos, and the rotation velocity increases until it reaches the velocity dispersion in the Local Group of $\sim 200$ \kms. 

We, here, encounter a question about "temperatures" of dark matter. Inside small galaxies, rotation velocities are as small as $\sim 100$ \kms, much less than those in giant spirals and intracluster space, and therefore, the potential is shallow. Accordingly, the dark matter is cooler, so that it is gravitationally bound to the shallow potential. On the other hand, the halo continuously merges with halos of bigger galaxies and/or intracluster matter with higher velocities of $\sim 200$ \kms. Thus, the temperature of dark matter increases from inside to outside of a dwarf galaxy. 
This temperature transfer of dark matter occurs also from inside the outskirts of the Galaxy at  $R\sim 100-150$ kpc, where the rotation velocity is $\sim 100$ \kms according to the well fitted NFW model, to outside where the Local Group dark matter with velocity dispersion of $\sim 200$ \kms is dominant.

We may consider that there are three components of dark matter. First, the galactic dark matter which defines the mass distribution in a galaxy controlling the outer rotation curve; second, extended dark matter filling the entire Local Group having a velocity dispersion as high as $\sim 200$ \kms, which gravitationally stabilize the Local Group; and finally, uniform dark matter having much higher velocities originating from supergalactic structures. The third component, however, does not significantly affect the structure and dynamics of the present Local Group. We may therefore speculate that at any place in the Galaxy, there are three different components of dark matter having different velocities or different temperatures. They may behave almost independently from each other, but are interacting by their gravity. It would be an interesting subject for the future to perform a simulation along the line of Sawa and Fujimoto (1998), but taking into account the different dark matter components of dark halos and the Local Group's dark mass.


\begin{thebibliography}{}
\def\r{\bibitem[]{}} 

\r Begeman, K. G., Broeils, A. H., Sanders, R. H. 1991, MNRAS, 249, 523 

\bibitem[Blitz et al.(1982)]{1982ApJS...49..183B} Blitz, L., Fich, M., \& Stark, A.~A.\ 1982, \apjs, 49, 183 

\r Burkert, A., 1995, ApJ, 447, L25.  

\r Burton, W. B. , Gordon, M. A. 1978 AA 63, 7.

\r Carignan, C., Chemin, L., Huchtmeier, W. K., Lockman, F. J. 2006 ApJ 641 L109 
 
\r Clemens, D. P. 1985. {\it Ap. J.} 295:422 
 
\r Corbelli, E., Salucci, P. MNRAS 2000, 311, 441. 

\r Cox, T. J., Loeb, A. 2008 MNRAS 386, 461

\bibitem[de Vaucouleurs(1958)]{1958ApJ...128..465D} de Vaucouleurs, G.\ 1958, \apj, 128, 465 

\r Demers, S., Battinelli, P. 2007 AA, 473, 143.

\r Einasto, J.; Kaasik, A.; Saar, E. 1974 Nature, 250, 309 


\bibitem[Fich et al.(1989)]{1989ApJ...342..272F} Fich, M., Blitz, L., \& Stark, A.~A.\ 1989, \apj, 342, 272 

\r Fich, M,, Tremaine, S. 1991 ARAA, 29, 409. 

\bibitem[Freeman(1970)]{1970ApJ...160..811F} Freeman, K.~C.\ 1970, \apj, 
160, 811 

\r Gentile, G., Salucci, P., Klein, U., Granato, G. L. 2007 MNRAS 375, 199. 

\r Hayashi, H., Chiba, M. 2006, PASJ 58, 835. 

\r Honma, M., Bushimata, T., Choi, Y. K.; Hirota, T., Imai, H. et al. 2007 PASJ 59, 839.

\r Honma, M., Sofue, Y., 1997, PASJ 49, 453 

\r Jimenez, R., Verde, L., Oh, S. P. 2003 MNRAS 339, 243. 


\r Kahn F.D., Woltjer L., 1959, ApJ, 130, 705

\r Kochanek, C. 1996 ApJ 457, 228. 

\r Kopylov, A. I., Tikhonov, N. A., Fabrika, S., Drozdovsky, I., Valeev, A. F. 2008 MNRAS 387L, 45.

\r Kulessa, A S.; Lynden-Bell, D.  1992 MNRAS 255, 105 

\r Li Y.-S., White S. D. M.,  2008 MNRAS 384, 1459.

\r Loeb A., Reid M. J., Brunthaler A., Falcke H., 2005, ApJ, 633, 894

\r Mateo, M. L. 1998, ARAA, 36, 435

\r Navarro, J. F., Frenk, C. S., White, S. D. M., 1996, ApJ, 462, 563 

\r Peebles, P. J. E.; Phelps, S. D., Shaya, E. J.; Tully, R. B. 2001 ApJ, 554, 104 
 
\r Peebles P. J. E., Phelps S. D., Shaya E. J., Tully R. B., 2001, ApJ, 554, 104
\r Prada, F., Vitvitska, M., Klypin, A., Holtzman, J. A., Schlegel, D. J., et al. 2003, ApJ, 598, 260
 
\r Seljak, U. 2002 NNRAS 334, 797.

\r Sawa, T., Fujimoto, M., PASJ 2005, 57, 429. 

\r Sofue, Y., Honma, M., Omodaka, T. 2008 PASJ, to appear.
 
\r Sofue, Y., Rubin, V. C. 2001 ARAA 39, 137  

\bibitem[Spergel et al.(2003)]{2003ApJS..148..175S} Spergel, D.~N., et al.\ 
2003, \apjs, 148, 175 

\r Valtonen M. J., Byrd G. G., McCall M. L., Innanen K. A., 1993, AJ, 105, 886.

\r van den Bosch, F. C., Robertson, B. E., Dalcanton, J. J., de Blok, W. J. G 2000 AJ 119, 1579. 

\r van der Marel R. P., Guhathakurta P., 2008 ApJ, 678, 187 
 
\r  Zaritsky, D., Olszewski, E W., Schommer, R A., Peterson, R C., Aaronson, M 1989 ApJ 345, 759 

\r Zaritsky D., Smith R., Frenk C., White S.D.M., 1997, ApJ, 478, 39

\end{thebibliography}
\end{document}